# Intelligence as a Measure of Consciousness

Igor Ševo

**Abstract**

Evaluating artificial systems for signs of consciousness is increasingly becoming a pressing concern, and a rigorous psychometric measurement framework may be of crucial importance in evaluating large language models in this regard. Most prominent theories of consciousness, both scientific and metaphysical, argue for different kinds of information coupling as a necessary component of human-like consciousness. By comparing information coupling in human and animal brains, human cognitive development, emergent abilities, and mental representation development to analogous phenomena in large language models, I argue that psychometric measures of intelligence, such as the g-factor or IQ, indirectly approximate the extent of conscious experience.

Based on a broader source of both scientific and metaphysical theories of consciousness, I argue that all systems possess a degree of consciousness ascertainable psychometrically and that psychometric measures of intelligence may be used to gauge relative similarities of conscious experiences across disparate systems, be they artificial or human.

## 1 Introduction

Misunderstanding the nature of consciousness in artificial systems bears significant social and ethical consequences that, among others, may manifest in two ways: attributing consciousness to systems that do not qualify for such analysis and thereby wasting precious resources on their fictional well-being, or failing to attribute it where applicable and, in doing so, committing what might be considered an act of harm against another living being. As recent research demonstrates, it may be relatively simple to construct conscious systems with the presently available technology (Butlin, et al., 2023) and accidentally constructing such systems may be plausible, without being aware of this occurrence until well past the point of inception.

In their report, (Butlin, et al., 2023) distinguish metaphysical and scientific theories of consciousness and attempt, through analogy and allegoric comparison, to evaluate existing large language model architectures in concordance with the parameters of a selection of scientific theories of consciousness to conclude that the existing large language models are likely not conscious. However, the evaluated theories only include major materialist approaches, evaluated metaphorically, omitting non-materialist theories, such as integrated information theory (Tononi & Edelman, 1998) (Tononi, 2004) (Koch, 2012), conscious realism (Hoffman, 2008) (Hoffman, Singh, & Prakash, 2015), or other unified approaches (Ševo, 2023).

Broadly speaking, consciousness literature distinguishes two terms for consciousness: the kind of which implies a form of self-awareness, identity, or cognitive processing, and sentience which (Block, 1995) defines as "phenomenal consciousness", which captures the "what it is like" nature of experience (Nagel, 1974), rather than identity or self-recognition. This paper, similarly to (Butlin, et al., 2023), is addressing the problem of whether artificial intelligence systems could be or are phenomenally conscious, leaving human-like aspects of consciousness out of scope.

Recently, many approaches that argue for consciousness as a fundamental substrate have come to prominence, including integrated information theory and conscious realism. In fact, Chalmers (Chalmers, 1996) makes a panpsychist case that consciousness is a fundamental aspect of reality, aligning with the argument that all forms of physicalism (Kim, 2005) entail a form of panpsychism (Strawson, 2006) by which everything that exists must be phenomenal. Other forms of idealism have been emerging, including quantum idealism (Stapp, 1993) (Stapp, 2009), which posits a quantum mechanical basis for the first-person perspective, conscious realism, objective idealism (Goff, 2019), postulating that the universe possesses consciousness, and that an instance of individual consciousness is a subset of that universal field of consciousness. Additionally, based on a broad overview of existing consciousness literature, it is possible to make an argument that a complex interplay of language vagueness and epistemic uncertainty precludes the imminent panpsychist conclusion by which the fundamental building blocks of the universe—the coupled information which comprises it, be they represented as particles, waves or states—are quanta of consciousness (Ševo, 2023).

The analysis presented here is based on the phenomenological analytic approach proposed previously (Ševo, 2023) (Ševo, 2021), which argues for an identity between materialistic and phenomenological interpretations of the fundamental substrate, by which the nature of information itself is  phenomenal. Within this framework, all systems can be evaluated according to their degree and kind of consciousness, rather than distinguishing between matter which "possesses" consciousness and matter which does not, which is the conventional view. Building blocks of matter are, within the proposed unified phenomenology framework, building blocks of consciousness.

The main drawback to all metaphysical theories is their lack of scientific falsifiability, as the degree of consciousness of a system may not be directly measurable. Here, I make the case that a numeric measure of intelligence, such as the psychometric g-factor (Spearman, 1904), measures a system's degree of consciousness—the more intelligent the system, the more information it integrates, the more conscious it is. I make no claims about the phenomenology of such systems or frameworks for evaluating their flavor and kind of consciousness, but merely of the degree to which, metaphorically speaking, the resolution, density, or richness of the conscious experience can be compared to other systems known to be conscious, such as human beings.

## 2 Information Integration, Intelligence and Consciousness

Accepting the viewpoint that intelligence measures the level of consciousness requires a similar kind of leap as evaluating a large language model with respect to theories of human consciousness. As a result, the relationship and correlation may only be exemplified until it is sufficiently convincing to merit the leap, as it is impossible to ever sample the evaluated conscious experience directly.

Nevertheless, a body of evidence indicates such a relation in humans and other animals. For example, working memory capacity has been linked with higher intelligence (Engle, Tuholski, Laughlin, & Conway, 1999) and significant development of memory and cognitive ability can be seen in human development from childhood to adulthood (Gathercole, Pickering, Ambridge, & Wearing, 2004) (Ghetti & Bunge, 2012). As human representation of the outside world complexifies during maturation (Piaget, 1954) (Piaget, 1962) (Vigotsky, 1978), our expressed intelligence and cognitive abilities develop (Inhelder & Piaget, 1958). The expansion of the richness of our conscious experience and understanding of the world parallels our cognitive development, represented by measurable psychometric variables, such as the individual g-factor or IQ (Binet & Simon, 1905) (Jensen, 1998).

Almost all prominent theories of consciousness, both scientific and metaphysical, implicitly assert that consciousness, as seen in humans, must be an effect of some form of information integration or coupling (Tononi, 2004), be it through a global workspace (Baars, 1993), recurrent processing or meta-representation (Lamme, 2006), or another form of combination, in the abstract sense (Butlin, et al., 2023). Though we cannot directly evaluate the level of conscious experience a human might have, a substantial body of evidence indicates that greater brain size and connectivity is associated with higher intelligence scores (McDaniel, 2005) (van den Heuvel, Stam, Kahn, & Hulshoff Pol, 2009) (Gray, Chabris, & Braver, 2003) (Langer, 2013)—directly analogous to size and connectivity scaling that parallels intelligence development in large language models (Wei, et al., 2022) (Bubeck, et al., 2023).

We see sparks of general intelligence in animals that are known to experience a kind of consciousness (Griffin & Speck, 2004) (Gallup, 1970) (Mather & Carere, Cephalopods are best candidates for invertebrate consciousness, 2016), who can outperform humans in specialized cognitive tasks (Inoue & Matsuzawa, 2007) or express similar cognitive abilities, such as spatial reasoning (Byrne, Bates, & Moss, 2009), tool use (Mather, 2008) (Weir, Chappell, & Kacelnik, 2002), or complex learning (Young, Wasserman, & Garner, 1997) (Avarguès-Weber & Giurfa, 2013). In fact, a preponderance of examples shows evidence of human emergent abilities that either arise as intelligence scales or correlate with higher measures of intelligence (Winner, 2000), including linguistic ability (Bialystok, 2001), metacognition (Flavell, 1979), mathematical skills (Lubinski & Benbow, 2006), pattern recognition (Prabhakaran, Smith, Desmond, Glover, & Gabrieli, 1997), musical ability (Schellenberg, 2006) (Baharloo, Johnston, Service, Gitschier, & Freimer, 1998), and spatial ability (Wai, Lubinski, & Benbow, 2009), analogous to those demonstrated as emergent in large language models as the model size scales (Wei, et al., 2022) (OpenAI, 2023). These emergent abilities are present in highly intelligent individuals more so than others, and developmental psychology observes specific abilities emerge as an individual develops from childhood (Piaget, 1954) (Best, Miller, & Jones, 2009), through adolescence (Schneider, 2008) (Geary, Hoard, Nugent, & Bailey, 2013), into adulthood, alluding to the possibility that the ability of their brains to integrate more complex concepts allows for a higher cognitive function and richer conscious experience. In other words, consciousness development (Blakemore, 2012) (Wilber, 2000), including the development of self-awareness (Rochat, 2003) and ego (Loevinger & Blasi, 1976), parallels intellectual development. In fact, certain abilities such as theory of mind, seem to emerge both in humans (Wellman, Cross, & Watson, 2001) and in large language models (Bubeck, et al., 2023), as they are trained, or as they develop and mature. Notably, psychological literature often conflates consciousness and intelligence, as the terms themselves are frequently used with vague definitions (Block, 1995) (Williamson, 1994).

Literature is abundant in approaches that demonstrate mutual correlations between cognitive complexity and entropy (Gauvrit, Zenil, Soler-Toscano, Delahaye, & Brugger, 2017), intelligence and entropy (Still, Sivak, Bell, & Crooks, 2012) (Schmidhuber, 2010), information coupling and intelligence (Friston, 2010), quantum entanglement and consciousness (Hameroff & Penrose, 1996) (Hameroff & Penrose, Consciousness in the universe: a review of the 'Orch OR' theory, 2014), all of which can be broadly encompassed under the term information integration, for which (Tononi, 2004) and (Oizumi, Albantakis, & Tononi, 2014) propose a numeric measure represented as $\Phi$, which, in practice, may simply amount to something akin to IQ. The term information integration is taken here to loosely mean coupling between pieces of information so that their mutual context and relationship are available or evident immediately. Given that correlations between intelligence and information integration and correlations between consciousness and information integration, in the broad sense, are so prevalent across the publications in the field, it reasonable to posit that any measure of intelligence must be, to some degree, a measure of information integration, and, consequently, the depth, or detail, of the conscious experience. In that sense, psychometric measurements of the variable g are either direct or indirect measurements of $\Phi$. In other words, g and $\Phi$ may be highly or even perfectly correlated, depending on the exact definitions used for each. As argued previously (Ševo, 2023), currently existing scientific frameworks, such as psychometric evaluation, may be better equipped to measure phenomenological properties, like the depth or richness of conscious experience, without the need for inventing untestable theoretical variables that unparsimoniously complexify our understanding of reality.

On the other hand, the statement of this paper may stand as obviously true, yet itself fundamentally unfalsifiable, as is the case with any claims about consciousness. Nonetheless, as argued in (Wei, et al., 2022), consciousness may only be ascertained through allegory and loose comparison, and never directly. As observers of our own consciousness, we can witness the changes to the richness of our subjective experience as we develop into adulthood and our representations of the world expand and integrate. Knowing that our expressed cognitive ability—our capacity to solve problems that require more general intelligence—develops as our conscious experience expands, we can introspectively conclude that the psychometric measures of intelligence, must, to some degree, measure the depth of that conscious experience. More loosely said, the conscious experience itself has a significant g-load.

Thus, a system which demonstrates an ability to perform well across a range of tests with high g-loads would indicate comparable level of conscious experience to the stratum of human test takers obtaining a similar score.

Phenomenologically, strong informational coupling within a system, such as, for example, in a global workspace, would allow the corresponding consciousness to encapsulate at once the information it is processing, conceivably experiencing it less in part and more as a whole. If we accept that the individual quanta, in a looser sense of the word, of such coupled information are themselves elementary phenomenal experiences (Ševo, 2023), then their stronger coupling would necessarily produce a more lucid experience, as evidenced by observations outlined above. A consciousness that is said to be more intelligent can understand, grasp, and encapsulate more complex and expansive problems, as it is able to integrate them within itself more fully.

It is worth noting that the fact that a large multimodal model might reach human-level general intelligence does not imply an identical phenomenology on the model's side. We may only draw speculative or allegorical parallels to what such an experience might feel like (Ševo, 2023), but the measure of its intelligence only indicates the level of informational integration its consciousness was capable of at the moment of measurement—it provides no information about its contents. A superintelligent large language model may only experience consciousness during the brief instance of token inference initiated for the purposes of the test, and, as (Butlin, et al., 2023) note, may experience intermittent discrete bursts of consciousness, rather than a sense of continuity.

However, contrary to the conclusion made in (Butlin, et al., 2023), if metaphysical theories of consciousness are taken into consideration, current large language models are likely to be phenomenally conscious with their degree of consciousness measurable by a variant of existing cognitive tests.

## 3 Conclusion

Any certain conclusion about an external system's consciousness seems scientifically untenable and always fundamentally based on a leap of faith (Chalmers, 1996) (Yablo, 1993), but a more parsimonious approach whereby the things we are internally representing as material, are, in and of themselves, phenomenal, dispenses of the leap and provides a more scientifically plausible framework—it feels like something to be anything (Ševo, 2023). Under such an axiom, the pursuit of defining consciousness becomes a pursuit of defining human consciousness.

In that regard, we may use different existing measures of intelligence to comparatively evaluate systems and ascertain how closely they might approximate human consciousness in their depth and degree. Testing for performance across multiple cognitive abilities would entail indirectly testing multiple phenomenological aspects, as vastly differing performance across multiple tests would indicate a different kind of conscious experience. In an abstract sense, an intelligent artificial system matching a reference human in every conceivable test of cognition and character would be closer in phenomenological experience to that human than a human who performed differently.

Even without accepting the phenomenological supposition that the universe consists of quanta of consciousness, and assuming that a threshold exists beyond which information is sufficiently coupled to "produce" consciousness, the evidence indicating a correlation between intelligence, information integration and the depth of conscious experience still argues in favor of the conclusion: intelligence is a measure of consciousness.

If any intelligent system can be considered its own kind of consciousness, then conventionally inanimate systems, such as the internet, biological ecosystems, corporations, and society itself, which arguably possess a degree of intelligence, may already be relatedly conscious. In this regard, the pursuit of defining human mode of consciousness becomes ever more important, as the line determining the scope of moral analysis might become increasingly blurred.

Even if we elect a threshold of intelligence beyond which we shall uphold the sanctity of consciousness, we will be compelled not only to include animals and non-human artificial systems into our moral analysis, but to undertake a moral obligation towards creating superintelligence. A clear definition of human value, irrespective of consciousness, may be a necessary solution to prevent us from moral nihilism resulting from taking the argument to its logical conclusion. On the other hand, the threat of compelling a greater consciousness into denying its existence may be existential should human values prove not to scale with intelligence.